\documentclass[manuscript]{aastex}

\slugcomment{to appear in the Astrophysical Journal, September 2009}

\shorttitle{The California Molecular Cloud}
\shortauthors{Lada Lombardi \& Alves.}

\def\msun{M$_\odot$}
\def\msunsp{M$_\odot$ \ }
\def\cc{cm$^{-3}$}

\def\13co{$^{13}$CO}

\begin{document}


\title{The California Molecular Cloud}


\author{Charles J. Lada }
\affil{Harvard-Smithsonian Center for Astrophysics, 60 Garden Street
Cambridge, MA 02138}
\email{clada@cfa.harvard.edu}

\author{Marco Lombardi}
\affil{European Southern Observatory, Karl-Schwarzschild-Strasse 2, 85748, Garching Germany}
\email{mlombard@eso.org}

\and

\author{Jo\~ao F. Alves}
\affil{Calar Alto Observatory,  C/Jesus Durban Remon, 2-2, 04004 Almeria, Spain}
\email{jalves@caha.es}

\begin{abstract}
We present an analysis of wide-field infrared extinction maps of a region in Perseus just north of the Taurus-Auriga dark cloud complex. From this analysis we have identified a massive, nearby, but previously unrecognized, giant molecular cloud (GMC).  Both a uniform foreground star density and measurements of the cloud's velocity field from CO observations indicate that this cloud is likely a coherent structure at a single distance. From comparison of foreground star counts with Galactic models we derive a distance of 450 $\pm$ 23 parsecs to the cloud. At this distance the cloud extends over roughly 80 pc and has a mass of $\approx$ 10$^5$ \msun, rivaling the Orion (A) Molecular Cloud as the largest and most massive GMC in the solar neighborhood. Although surprisingly similar in mass and size to the more famous Orion Molecular Cloud (OMC) the newly recognized cloud displays significantly less star formation activity with more than an order of magnitude fewer Young Stellar Objects (YSOs) than found in the OMC, suggesting that both the level of star formation and perhaps the star formation rate in this cloud are an order of magnitude or more lower than in the OMC. Analysis of extinction maps of both clouds shows that the new cloud contains only 10\% the amount of high extinction (A$_K >$ 1.0 mag)  material as is found in the OMC. This, in turn, suggests that the level of star formation activity and perhaps the star formation rate in these two clouds may be directly proportional to the total amount of high extinction material and presumably high density gas within them and that there might be a density threshold for star formation on the order of n(H$_2$) $\approx$ a few $\times$ 10$^4$ \cc. 
\end{abstract}



\section{Introduction}

In this paper we report and analyze new observations of a little studied molecular cloud in Perseus. This cloud attracted our attention in the course of analyzing  an extinction mapping survey of the Taurus-Auriga-Perseus region. This survey was obtained as part of an ongoing systematic program to construct and analyze wide-field infrared extinction maps of prominent dark clouds in the solar vicinity. These clouds include the Pipe Nebula region \citep[LAL06]{LAL06}, the Ophiuchus and Lupus cloud clouds \citep{LLA08}, the Orion/MonR2 region \citep[LAL09]{LAL09b} as well as the Taurus-Perseus-Auriga region \citep[LLA09]{LAL09a}.  Examination of the latter survey revealed a cloud with significantly more foreground star contamination than either the Taurus-Auriga or Perseus clouds. Inspection of an early CO survey of the same region \citep{UT87} revealed the cloud to also have a distinctly larger radial velocity than either the Taurus or Perseus clouds. Using a modern version of the classical method pioneered by \citet{Wolf23}, we combined measurements of the foreground star density with galactic models to derive a distance to the cloud of 450 pc. This is considerably more distant than either the Taurus-Auriga (150 pc) or Perseus (240 pc) clouds but comparable to the more famous Orion molecular clouds (400 pc). We determine the mass of this complex to be approximately 10$^5$ \msunsp making it a bona fide giant molecular cloud (GMC). Interestingly this cloud is very similar in size and shape to the Orion A GMC, but somewhat more massive. Yet despite being perhaps the most massive GMC within 0.5 kpc of the solar system, surprisingly little is known about it.  

With this paper we begin to remedy this situation by providing measurements of some of its basic physical properties, including distance, size, mass, structure and star forming activity. The cloud lies entirely within the Perseus constellation and so to avoid confusion with the closer, better known (Perseus) cloud  to its south, we designate this cloud the California Molecular Cloud after the most prominent optical feature associated with it: the California Nebula. We compare the properties of the California Molecular Cloud to those of the Orion A Molecular Cloud  and find the striking similarities in masses, sizes and shapes of these two clouds to be in stark contrast to the striking differences in their star formation activity and density structure.

\section{Observational Data}

\subsection{Extinction Maps}

Infrared extinction maps of the California Molecular Cloud (CMC) were constructed using the NICER  and NICEST extinction mapping methods \citep{Lom05, Lom09} with infrared observations from the Two Micron All Sky Survey (2MASS; \citet{ketal94}). The maps were constructed as part of a wide field survey of the Taurus-Auriga-Perseus molecular clouds (LLA09) and their surroundings.  The detailed procedures used for creating the map are described in LLA09 and in LAL04. Briefly the infrared color excesses were derived for 23 million stars in a 3500 square degree area including the Taurus-Auriga, Perseus and California molecular clouds. The color excesses were derived by subtracting from the observed colors of each star the corresponding intrinsic colors derived from stars in a nearby control field with negligible extinction. These color indices were then scaled by the appropriate extinction law (Indebetouw et al. 2005) to derive infrared extinctions, (e.g., $A_{K}$) to each star.  The individual extinction measurements were then smoothed with a gaussian weighting function (FWHM = 80.0 arc sec) and sampled with a spatial frequency of 40.0 arc sec to produce a completely sampled grid from which the map of the California Molecular Cloud was constructed.  The resulting map produced by the NICER method is presented in Figure 1.

Besides the California Nebula located at its southern edge, the previously cataloged dark clouds 
L1441, L1442, L1449, L1456, L1459, L1473, L1478, L1482 and L1483  all appear to be part of the CMC complex. In addition the reflection nebula NGC 1579 which is illuminated by LK H$\alpha$ 101 is embedded at the eastern edge of the cloud.

\subsection{CO Maps}

CO maps were obtained from the archive of the Galactic plane survey performed with the 1.2 meter Millimeter-wave Telescope at the Center for Astrophysics (Dame, Hartmann \& Thaddeus 2001). The data were uniformly sampled with a spatial frequency equal to the  beamwidth of 8 arc minutes and with a velocity resolution of 0.65 km s$^{-1}$.  These maps represent a significant improvement over the undersampled maps of \citet{UT87}. Figure \ref{CO} shows the spatial map of $^{12}$CO emission from the CMC in Galactic coordinates and a position-velocity (pv) map through the cloud complex made along galactic longitude. The pv map has been integrated over 3 degrees in Galactic latitude as shown in the figure and roughly parallels the primary axis of the cloud.  The CO spatial map shows basically the same overall morphology as the extinction map despite the large (x~6) difference in angular resolutions between the maps. (This is perhaps most clearly evident by comparison with Figure 6 where the extinction map is plotted in Galactic Coordinates.) The pv map shows that the  cloud is continuous in velocity as well as spatially in projection and this provides strong evidence that the emission originates in a single contiguous cloud at the same distance. There is a significant velocity gradient along Galactic longitude between $l_{II} = $ 156 and 163 degrees.  The magnitude of this gradient is approximately 0.9 km s$^{-1}$ deg$^{-1}$ or roughly 0.1 km s$^{-1}$ pc$^{-1}$, which is typical, if not relatively modest, for a GMC.

\section{Basic Physical Parameters}

\subsection{Distance}

We derive the distance to this cloud using the density of foreground stars in a manner
similar to the classical method of Wolf (1923) and following the method used by
Lombardi et al. (2009a; 2009b) to determine the distances to the Taurus-Auriga, Perseus,
Orion and Mon R2 clouds. The Taurus-Auriga and Perseus clouds are in the same general direction of the Galaxy as the California cloud and the distances LLA09 derive for these clouds are in excellent agreement with VLBI parallax measurements.  

Briefly we first identify foreground stars in regions of high extinction. To do this we select all the high extinction pixels ($A_K >$ 0.6 magnitudes) in the map and search  for stars projected on these pixels that show ``no'' extinction, that is stars whose extinction is less than 3-$\sigma$ above the background. We then calculate the density of foreground stars takinig into account the area of the sky occupied by the high extinction pixels. In this way we found 119 foreground stars within an area of 0.27 square degrees yielding a foreground star density of $\rho$ $=$  440$\pm$ 30 deg$^{-2}$. The foreground density was also found to be uniform over the extent of the cloud indicating that the entire complex is likely at a single distance. We compared this density to the Galactic model  of \citet{rrdp03} which predicts stellar densities as a function of distance and direction in the Galaxy.  Figure \ref{dist} shows the plot of stellar density as a function of distance predicted from the Robin et al. model for the direction of the CMC. The observed foreground stellar density is also indicated and the intersection of this value with the model curve gives the distance to the obscuring dust cloud. This comparison yields a distance to the cloud of 450 $\pm$ 23 pc.

Previous distance estimates for this region range from 125 to 700 pc. \citet{e59} found two
extinction layers in this direction from optical photometry of field stars. The two layers
or clouds were found to have distances of 125 and 300-380 pc, respectively. These distances
are lower than our value but since the line-of-sight to the California cloud passes near both the Taurus-Auriga ( $\approx$ 140 pc distant) and the Perseus ($\approx$ 250 pc distant) clouds, the
layers identified by Ekl\"of are likely associated with these foreground complexes.   Recently
\citet{HAD04} derived a spectroscopic parallax distance of 700 $\pm$ 200 pc to 
the young stellar cluster embedded in NGC 1579 at the east end of the 
California cloud. This estimate is only marginally compatible with our star count estimate.
Finally, the Hipparcos parallax measurements \citep{petal97} of $\xi $ Per, the exciting star of the California Nebula, NGC 1499, suggests  a distance between 394  and 877 pc for the star,  consistent with both our distance and that of \citet{HAD04}.  However, for the remainder of this paper we adopt the star count distance of 450 pc for the cloud. As we show below, at this distance
the cloud rivals the Orion A (L1641) cloud as the largest and most massive GMC in 
the solar neighborhood (i.e., D $<$ 0.5 kpc).

\subsection{Mass, Size \& Structure}

We derive the cloud mass directly from the extinction map by integrating the dust column density over the area of the cloud and assuming a gas-to-dust ratio:

$$ M = D^2\mu\beta_K \int_{cloud} A_K(\theta)d^2\theta $$

\noindent
where $D$ is the distance to the cloud, $\mu$ is the mean molecular weight corrected for helium and $\beta_K$ is the gas-to-dust ratio [N(HI) $+$ 2N(H$_2$)]/$A_K$ =  1.67 x 10$^{22}$
cm$^{-2}$mag$^{-1}$ \citep{l55, betal78}.  The total mass of the CMC is found to be 1.12 $\times$ 10$^5$ \msunsp above an extinction of A$_K$ = 0.1 mag. This makes the mass of the CMC, comparable to, if not  slightly greater than, the mass of the Orion, L1641 GMC, usually considered to be the most massive molecular cloud within 0.5 kpc of the sun. 
 The CMC is characterized by a filamentary structure and extends over about 10 degrees on the sky which at the distance of  450 pc corresponds to a maximum physical extent of approximately 80 pc. Above an extinction of about  A$_K$ = 0.2 mag the cloud has a width of typically 1.5 degrees or 11 pc. However in regions of high extinction (A$_K > $1.0, mag) the cloud is extremely narrow, only barely resolved and less than roughly  0.2 pc in width (see Figure 1).  The northern portion of the cloud appears to split into 3 parallel filaments giving it the appearence of a trident (Figure 1). At the southern end the cloud again appears to split into (two) parallel filaments which are shorter and more closely spaced.

The distribution of mass within the CMC provides useful information about its internal structure
and physical state. Figure \ref{massfrac} shows the cummulative mass fraction of material in the cloud as a function of (infrared) extinction. This profile was generated using extinction  derived by the NICEST method which more accurately measures the highest extinction regions than using data generated with the NICER method. The function falls very steeply from low to high extinctions. For example, the cloud contains less than 1\% of its total mass at  extinctions of  A$_K  > $1 magnitude (i.e., approximatley A$_V$ $>$ 9 mag.) In Table 1 we list the masses of the cloud enclosed by increasing levels of extinction (1st column). As the extinction level  increases by a factor of 10 from A$_K$ = 0.1 to 1.0 magnitudes, the enclosed mass decreases by a factor of 100. Inspection of figure 1 shows that for the eastern half of the cloud (i.e., $\alpha_{2000} > 4^{\rm hr} 20^{\rm min}$) the highest extinction (A$_K$ $>$ 0.4 mag) regions are confined to a very narrow spine centered on the primary axis of the cloud. This indicates that the cloud is stratified, with an outwardly decreasing density gradient. This in turn implies that gravity is an important factor in determining the structure of the cloud. Such stratified structure is common for dense, dark cloud filamentary structures (e.g., \citet{allkp98, lal99}) and suggests that such clouds are dynamically evolved, quasi-stable objects in approximate pressure equilibrium with their surroundings. In this situation one expects gravity to be more important than turbulence in determining the structure of the cloud.

\subsection{Star Formation Activity}

Despite its large size and mass, the CMC appears to be very modest in its star formation
activity.  Besides the California Nebula, there are no prominent HII regions in this complex, indicating an absence of recently formed massive O stars. The California Nebula is excited by the O star, $\xi$ Per, but this star is a runaway O star from the Per OB2 association located more than 100 pc closer to the solar system and is not a product of star formation in the CMC. The best known and most prominent region of star formation within the cloud appears to be associated with NGC 1579, a reflection nebula containing a young embedded cluster with about 100 member stars \citep{aw08}.  The most massive young star in the CMC may very well be LK H$\alpha$ 101 which is a member of the embedded cluster in NGC 1579 and is likely an early B star \citep{HAD04}. 

To obtain an estimate of the overall star formation activity in the cloud we surveyed  IRAS point source catalog. We selected all IRAS sources within or near the boundaries of the cloud that had high quality fluxes in both the 25 and 60 micron bands. In all we found only 24 sources that satisfied our criteria and can be considered candidate YSOs. In figure \ref{iras} we display an image of IRAS emission from the CMC with the locations of the candidate YSOs indicated by crosses.   In Table 2 we list the IRAS YSO candidates we identified. All but two of these sources fall within the boundaries of the cloud delineated by the lowest contour in Figure 1. One of these is a late-type giant star (IRC 40094) and not a YSO (see table 2). Of the remaining sources, 17 are projected onto regions of highest extinction, which for the most part make up the very narrow filamentary spine of the cloud. In particular 11 sources line up along a dense and narrow filamentary ridge at the southern end of the cloud. Four of these sources exhibit the IRAS colors of Planetary Nebulae \citep{PM88}, but their close association with the high extinction ridge makes their status as Planetary Nebulae seem dubious. The reflection nebula NGC 1579 and its young cluster are embedded near the southernmost end of this ridge. The brightest IRAS source in the cloud is associated with LK H$\alpha$ 101, confirming that this source is likely the most luminous and massive star in the cloud.  Besides LK H$\alpha$ 101, five other candidates are associated with reflection nebulae and may be very young Class I protostars. Two of these are also associated with known HH objects and one of these may be an FU Ori star (see Table 2). The status of the remaining stars as YSOs requires further confirmation.   Overall the IRAS observations indicate that active star formation is occurring in this cloud, but at modest levels for a cloud this mass and size. 
  
\section{Analysis \& Discussion: Comparison with Orion}

In figure \ref{CMCOMC} we compare the NICER extinction images of the CMC and OMC at the same angular scale which closely corresponds to the same physical scale given the similarity of the distances to both clouds. The extinction map for the OMC was also derived using 2MASS data in a similar fashion to the CMC map (LAL09b). The CMC complex is easily as large as the OMC cloud. Moreover, the two clouds show a surprising similarity in their overall morphology. Both are filamentary in structure with a long central spine that appears to fork into two parallel filaments at both ends of the cloud.  The OMC does appear to have many more dark (high extinction) pixels than the CMC however. 

Despite the overall close similarity in size, filamentary structure, kinematics and mass, 
the two clouds differ dramatically in their level of star formation activity. For example, while LK H$\alpha$ 101 is the most massive and only known B star in the CMC, the Orion Nebula region of the OMC alone contains 20 OB stars, the most massive of which is an O5.5 star \citep{metal08}. Moreover,  the CMC appears to contain only one significant embedded cluster (associated with NGC 1579 and LK H$\alpha$ 101). This cluster contains about 100 members \citep{HAD04,aw08}. The OMC contains two significant embedded clusters (ONC  and L1641S) and numerous prominent groups or aggregates of YSOs (e.g., NGC 1977, OMC2,  L1641N, HBC 498, L 1641 C; \citet{pm08,ad08}). The ONC alone contains approximately 1700 members \citep{metal08, pm08} so is considerably (an order of magnitude) more populous than NGC 1579, while L 1641S is similarly rich as NGC 1579.  The L1641 dark cloud is the portion of the OMC south of the Orion Nebula and NGC 1999 regions. In this region of the cloud \citet{ctf93} used the co-added IRAS survey to identify about 100 sources likely to be YSOs. Recent reports of observations from the Spitzer Space Telescope suggest there may be as many as 750 YSOs in the L1641 portion of the cloud \citep{ad08} suggesting that the entire OMC presently contains 2000 - 2400 YSOs.  Clearly by any measure the star formation activity in the OMC dwarfs that in the CMC. Indeed, it is likely that the OMC has produced more than an order of magnitude more stars than the CMC. 

That two such similar nearby GMCs could have such drastically different levels of star formation is interesting. This indicates that although GMCs are always sites of star formation, the level of star formation can vary considerably and is not necessarily sensitive to the mass and size of the cloud.
This raises the interesting question about what factors determine the amount and rate of star formation in GMCs. Among the factors that could be important are time (evolution), structure and external influences. 

From the HR diagram for NGC 1579 constructed by \citet{HAD04} we estimate that the age of the clusters members would be 1-2 Myr for a distance of 450 pc, not all that different from the ages of the other nearby active star forming regions, including the ONC, Taurus, Perseus and Ophiuchus.  So any possible age differences between the CMC and the OMC are relatively small compared to the large difference in star formation activity between these two clouds. So it is unlikely that the difference in the yield of star formation between the two clouds is due to an age difference. Indeed, given the similar ages of the young stellar populations in both clouds, the enormous difference in the star formation yields indicates that star formation rates between the two clouds also significantly differ.

Although structurally similar in most respects there is one aspect that is significantly different between the two clouds.  Specifically, as mentioned above, the Orion cloud has more pixels at high extinction than does the California cloud. To make a more careful comparison of this difference in cloud structure we need to compare the NICEST extinction measurements of the two clouds at the same physical resolution and with the same foreground and background stellar densities.  
To do this we followed the prescription of Lombardi et al. (2009a), i.e. we recomputed a new, modified map for the California cloud using parameters that make the physical properties of the California and Orion map equivalent.  More precisely:
\begin{itemize}
\item We enlarged the pixel scale in the California cloud in order to have the same physical resolution of the Orion map, $0.113 \mbox{ pc pixel}^{-1}$;
\item The previous operation increased the already relatively large number of background stars per pixel used to build the California map.  Therefore, in order to have the same average number of background stars per pixel in both maps, we randomly discarded $\sim 40\%$ of the stars in the California field.
\item We estimated the density of foreground stars in both clouds, and we required both clouds to have the same number of foreground stars per pixel.  For this purpose, we added to the California cloud a few ($\sim 104 \mbox{ stars deg}^{-2}$) artificial foreground stars, generated from the colors observed in the control field.
\end{itemize}
Finally, we used this modified (lower resolution) map of California cloud to re-compute the cloud cumulative mass fraction as a function of extinction to enable a direct comparison with the corresponding profile derived for the Orion cloud. 

In Figure \ref{extprofile} we plot the cumulative mass fraction of each cloud as a function of infrared extinction ($A_K$). The two profiles are strikingly different. The CMC has a substantially lower fraction of its mass at high extinction. This difference is also tabulated in table 1, where
we have calculated the masses for both clouds above extinction levels, $A_K$,  of 0.1, 0.2 and 1.0 magnitudes. As can be ascertained from both the figure and table, the CMC contains somewhat less than 1\% of its mass above $A_K =$ 1 magnitude ($A_V \approx$ 9 mag) while the OMC contains slightly under 10\% of its mass above the same extinction level.    

This difference may provide a significant clue concerning the physical origin of the different levels of  star formation activity between the two clouds.  It has been known for some time that stars exclusively form in dense cores within GMCs \citep{eal92}. These cores have mean densities of typically 10$^4$ to 10$^5$ cm$^{-3}$. Thus one expects that the amount of star formation in a cloud will be directly related to the total amount of mass it contains at such high densities.  Star forming cores are also dark, characterized by visual extinctions ($A_V$) typically in excess of 5-10  magnitudes. Thus regions of relatively high density are also the regions of relatively
high extinction and high extinction can be used as a proxy for high density. For example, as pointed out earlier, regions with visual extinctions in excess of about 10 magnitudes are charactrized by size scales between 0.1 - 0.2 pc which corresponds to mean molecular hydrogen densities n$(H_2) > 1-2 \times 10^4$ cm$^{-3}$. It is interesting in this context to note that in the Ophiuchus cloud, such dense cores appear visible in submillimeter dust emission only where the mean extinctions are in excess of at least 7 visual magnitudes (Johnstone, DiFrancesco \& Kirk 2004). If we use A$_K$ = 1.0 mag as the indicator of the high extinction star forming material in a cloud, then we see that the amount of such material in the CMC is about an order of magnitude less than that in the OMC. The difference in cloud masses above this extinction threshold is of the same order as the difference in the number of YSOs in each cloud and perhaps even the star formation rate. Interestingly the CMC has roughly the same fraction of mass above this threshold as does the extremely quiescent Pipe Nebula (LAL06). In the Pipe the total mass above this threshold is only about 200 \msunsp since the Pipe is overall a much smaller complex. A recent deep Spitzer Space Telescope survey for YSOs in the Pipe uncovered only 18 such objects over the entire cloud (Forbrich et al. 2009). The ratio of known YSOs in the Pipe to the number of YSOs ($\sim$200) that we crudely estimate for the CMC, is roughly equal to 0.1 and is close in value to  the ratio (0.2) of the total mass of high extinction/ high density material in the Pipe to that in the CMC. 

If such trends hold with improved inventories of star formation activity in the CMC and with comparisons between additional clouds, this would imply that there exists a threshold extinction and presumably volume density for star formation and once reached there is a more or less constant star formation efficiency achieved  in the dense gas component of molecular clouds. We then expect the star formation rate to go as SFR $\sim$ $M_{dg} / \tau_{sf}$, where $M_{dg}$ is the total amount of dense gas (above the threshold) and $\tau_{sf}$ is the appropriate star formation time scale. If this time scale is given by the free-fall time ( $\tau_{ff} \sim (G\rho)^{-0.5}$) at the threshold density, then the star formation rate will be directly proportional to the total mass at high density. It has not escaped our notice that in external galaxies the comparison of global FIR star formation rates and molecular emission from the HCN molecule, a dense gas tracer, suggests such a relation may characterize the global star formation in galaxies ranging from normal spirals to ultraluminous starbursts \citep{gs04}. Indeed, this relation between star formation
rate and HCN luminosity appears to extend down to Galactic GMCs (Wu et al. 2005).

It is not clear why the CMC and OMC differ so significantly in their contents of high extinction/high density material. One possibility could be a difference in external environments. The OMC is associated with an OB association within which multiple supernovae are believed to have occurred over the last 10$^7$ years. As a result of the collective action of these supernovae the cloud may have been compressed by direct interaction with the supernovae remnants or as a result of increased pressure in the surrounding environment due to the hot bubble created by the supernovae.  The presence of Barnard's Loop and the Eridanus superbubble in the region immediately surrounding the OMC (Bally 2008) indicates the presence of a hot (10$^{5-6}$ K) and possibly high pressure medium external to the OMC. Similar activity is not observed near the CMC. However, whether this or some other factor is the cause of the differences in the extinction profiles between the two clouds cannot be presently ascertained with any confidence and requires further study.

\section{Summary \& Conclusions}

From an analysis of wide-field infrared extinction maps we have identified a nearby, previously unrecognized, massive molecular cloud  within Perseus.  Both a uniform foreground star density and measurements of the cloud's velocity field from CO observations indicate that the cloud is likely a coherent structure at a single distance. We designate this cloud the California Molecular Cloud due to its physical association with the well known California Nebula which is located on the cloud's southern border. From comparison of foreground star counts with Galactic models we derive a distance of 450 $\pm$ 23 parsecs to the cloud. At this distance the cloud extends over roughly 80 pc and has a mass (derived from the extinction measuremens) of $\approx$ 10$^5$ \msun.  The cloud thus rivals the Orion A Molecular Cloud as the largest and most massive GMC in the solar neighborhood.

Although surpisingly similar in mass and size to the more famous Orion Molecular Cloud the California Molecular Cloud displays significantly less star formation activity. There are more than an order of magnitude fewer YSOs in the CMC than in the OMC suggesting that both the level of star formation and perhaps the star formation rate in the CMC are correspondingly an order of magnitude or more lower than in the OMC. 

Analysis of extinction maps of both clouds shows that the CMC contains only about 10\% the amount of high extinction (A$_K >$ 1.0 mag)  material as is found in the OMC. This in turn suggests that the level of star formation activity and the perhaps star formation rate in these two clouds may be directly proportional to the total amount of high extinction material and presumably high density gas within the clouds and that there might be a density threshold for star formation of order n(H$_2$) $\approx$ a few $\times$ 10$^4$ \cc. Once this threshold is reached there would be a more or less constant star formation efficiency achieved in the dense gas. What adds somewhat to this surmise is that the CMC contains an order of magnitude more YSOs than are found in the extremely quiescent Pipe Molecular Cloud and, as it turns out, the Pipe Cloud contains only 20\% the amount of high extinction material as does the CMC (LAL06). Thus, the overall levels of star formation activity in these three clouds, the OMC, the CMC and the Pipe appear to be directly related to the total content of the high extinction/dense material within them. 

Finally, we find the structure of eastern half of the CMC to be well behaved and centrally condensed with the highest extinction material confined to a very thin spine along the primary axis of the elongated cloud. This systematically stratified structure suggests that this portion of the cloud may be in near pressure equilibrium with its surroundings and strongly self-gravitating.
It is also the region of most active star formation in the cloud.




\acknowledgments

We are grateful to Tom Dame for kindly providing the CO data and Figure \ref{CO}.

\clearpage



\begin{figure}
\epsscale{1.25}
\vskip 0.6in
\hskip -0.6in
\plotone{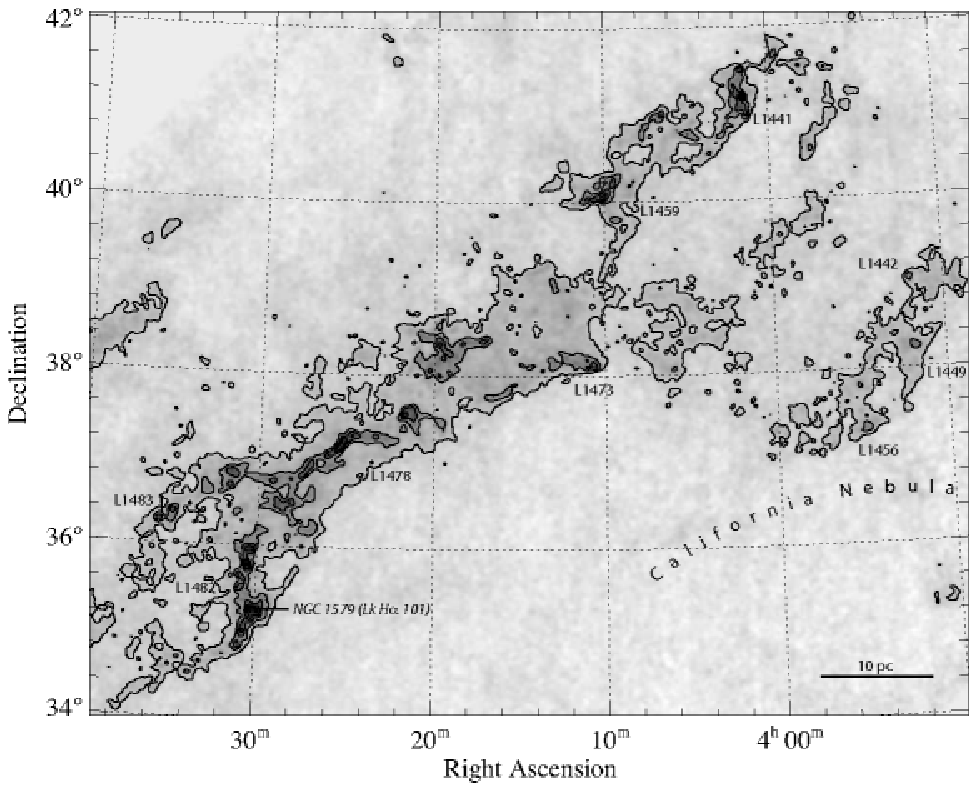}
\vskip -2.5in
\caption{The NICER Infrared extinction map of the California Molecular Cloud.  Contours start at an infrared extinction of A$_K$ = 0.2 mag and increase uniformly in steps of 0.2 mag. The locations of prominent Lynds dark clouds are indicated as well as the location of NGC 1579, a reflection nebula illuminated by the early B star, LKH$\alpha$ 101. The approximate location of the famous California Nebula is also indicated.\label{fig1}}
\end{figure}

\clearpage

\begin{figure}
\epsscale{1.00}
\plotone{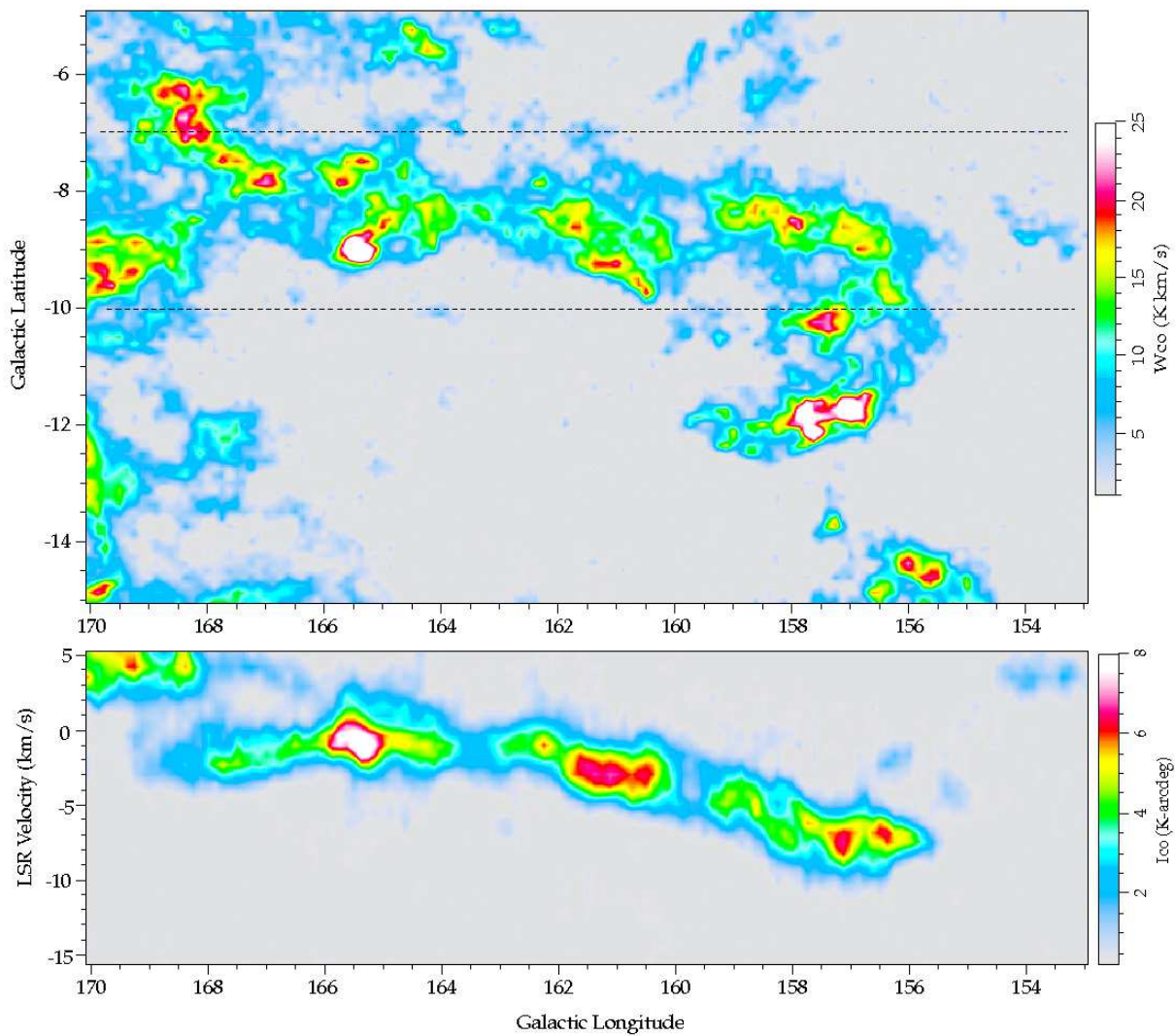}
\caption{Map of the integrated intensity of $^{12}$CO emission toward the California cloud from the survey of \citet{detal01}. Top panel shows the CO spatial map in Galactic coordinates. This map covers  roughly the same region as the extinction map in figure 1. (See also fig \ref{CMCOMC}). The bottom panel shows the position-velocity map along Galactic longitude. The parallel dashed lines in the top panel indicate the range of Galactic latitude integrated to produce the position-velocity map in the bottom panel.  \label{CO}}
\end{figure}

\clearpage

\begin{figure}
\plotone{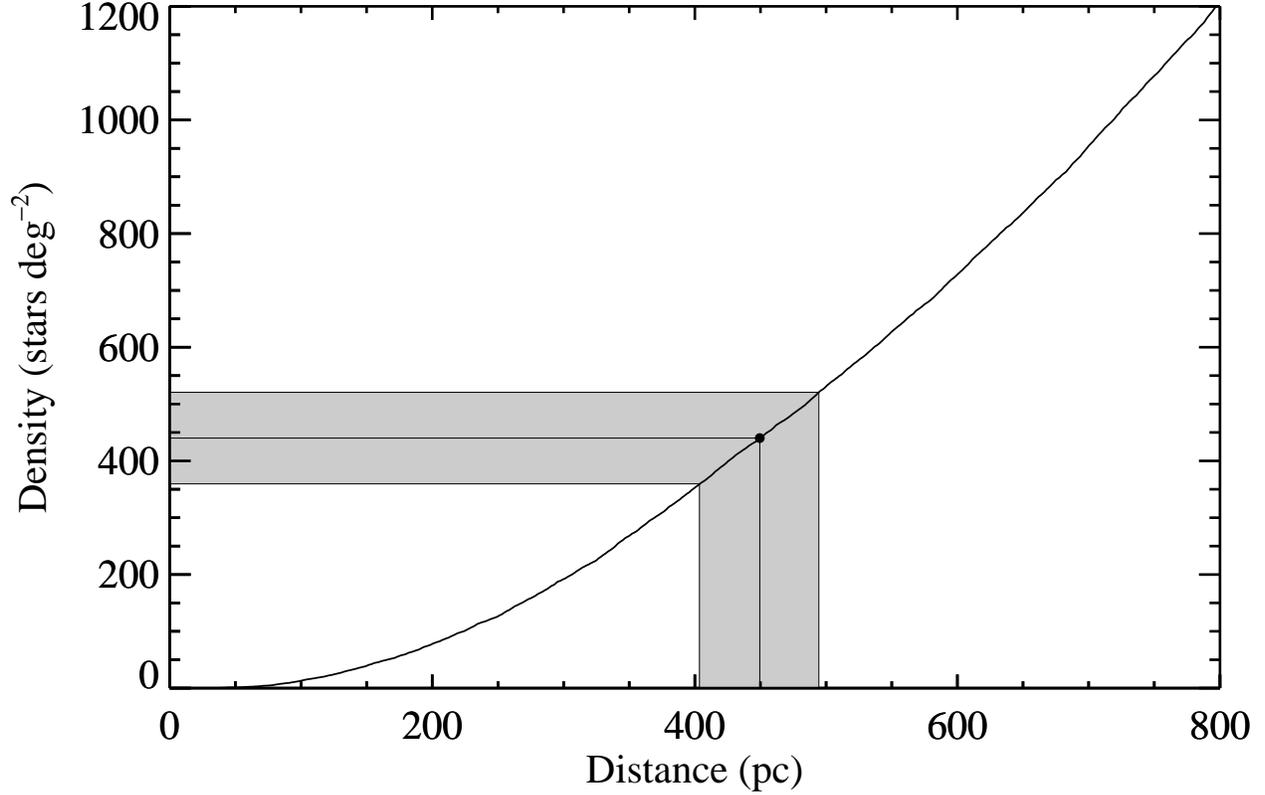}
\caption{The plot of foreground stellar density vs distance from the Galactic
models of Robin et al. (2003). The observed foreground star density and its 
uncertainties toward the highest extinction regions of the cloud are indicated by 
the horizontal lines. The intersection of these lines with the model give the
distance to the cloud and the corresponding uncertainties in that distance.  \label{dist}}
\end{figure}

\clearpage

\begin{figure}
\plotone{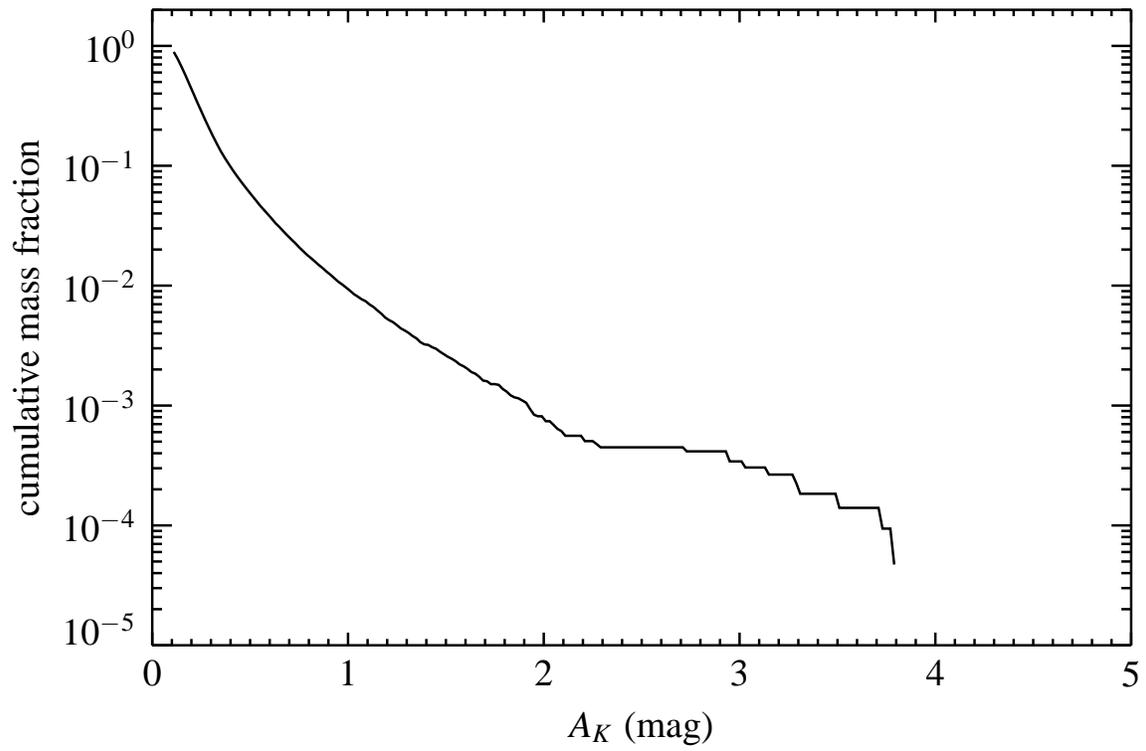}
\caption{ Plot of the cumulative mass fraction as a function of infrared extinction, A$_K$, for the CMC. \label{massfrac}}
\end{figure}

\clearpage

\begin{figure}
\plotone{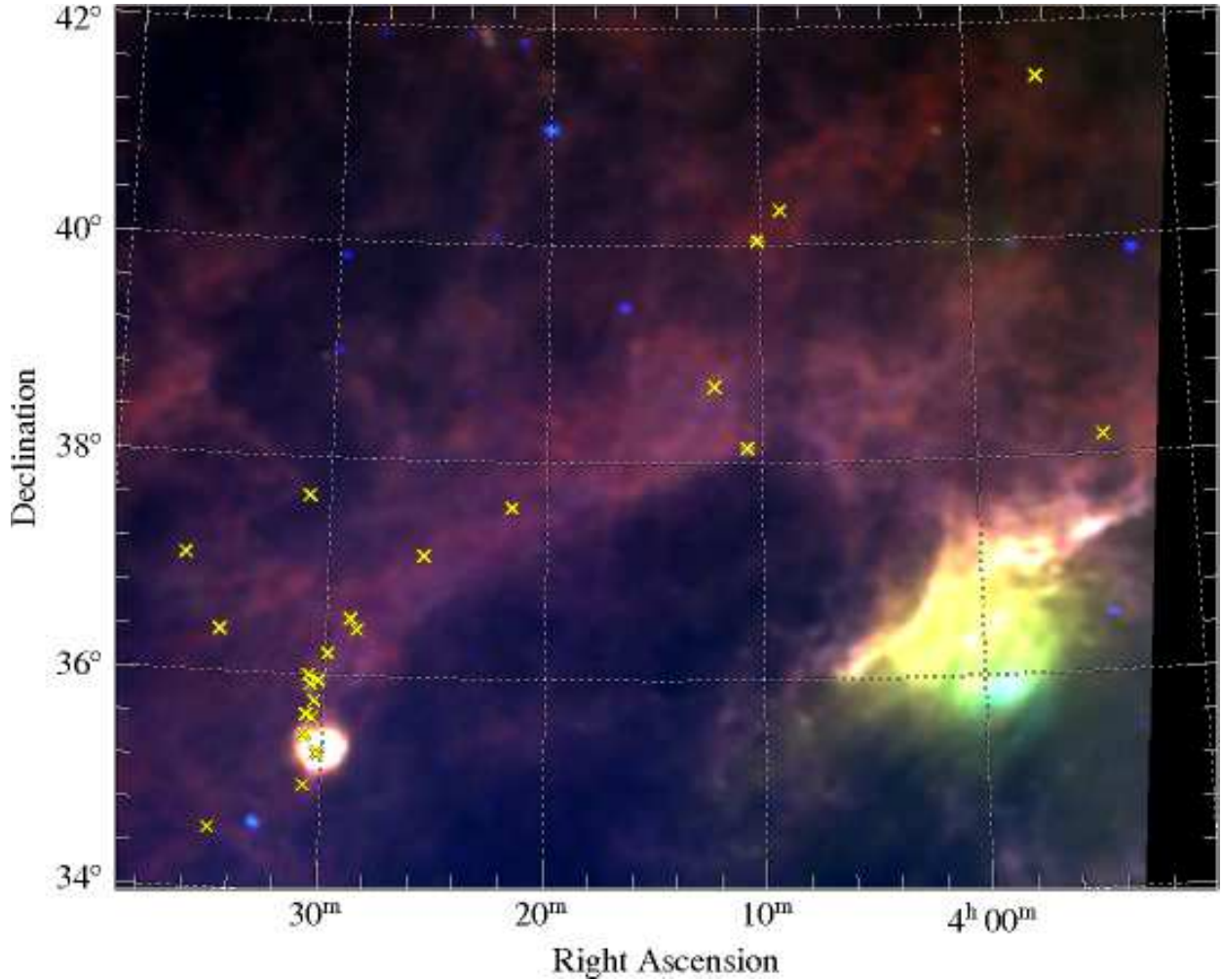}
\caption{Three color image of IRAS emission from the California Molecular Cloud. The region is the same as displayed in figure 1. Yellow crosses mark the positions of IRAS sources that are candidate YSOs. The bright, saturated nebulosity in the southwest is the California Nebula,
the compact bright saturated nebulosity in the southeast marks the position of NGC 1579 and
the young emission-line star LK H$\alpha$ 101.  \label{iras}}
\end{figure}

\clearpage

\begin{figure}
\vskip -0.3in
\hskip -0.8in
\epsscale{0.8}
\plotone{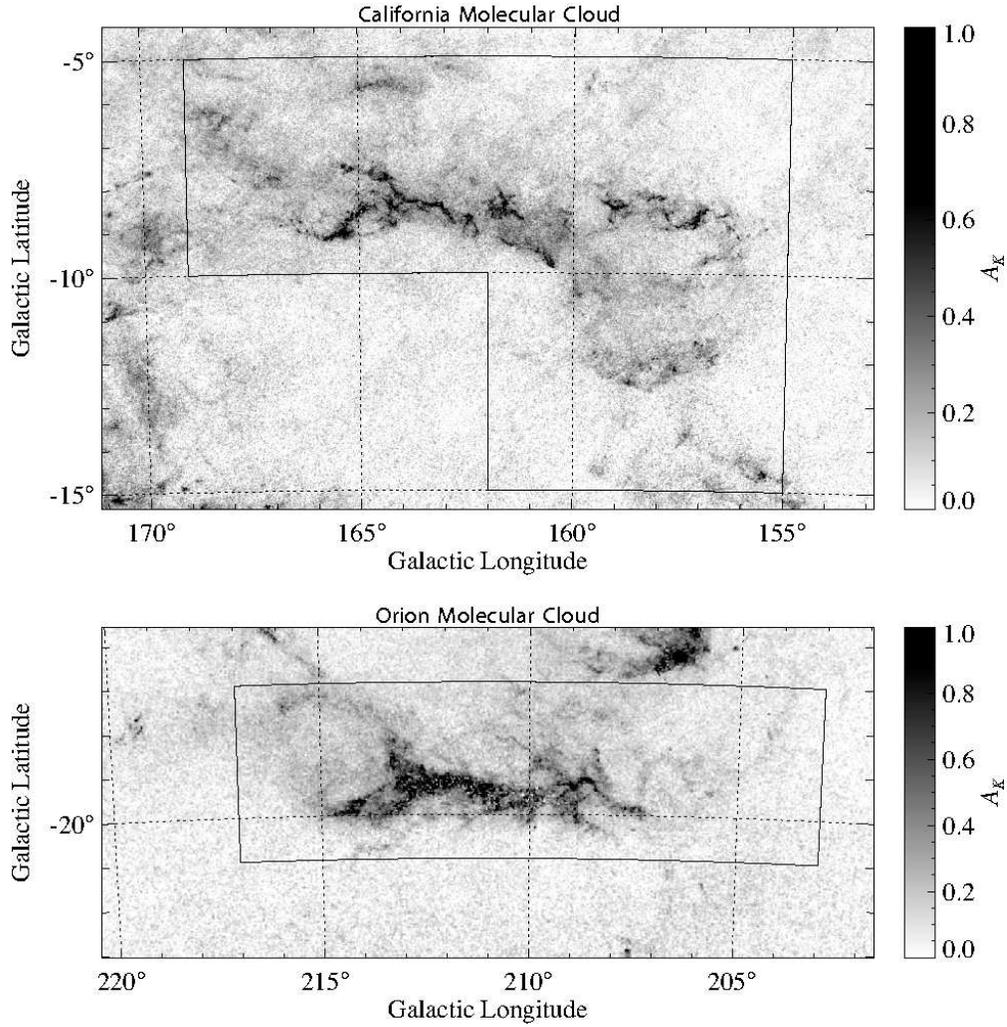}
\caption{Comparison of the infrared extinction images of the CMC and OMC clouds.
Given the similar distances, the CMC cloud is comparable in physical size and is very 
similar in structure compared to the better known OMC.  \label{CMCOMC}}

\end{figure}

\begin{figure}
\epsscale{.80}
\plotone{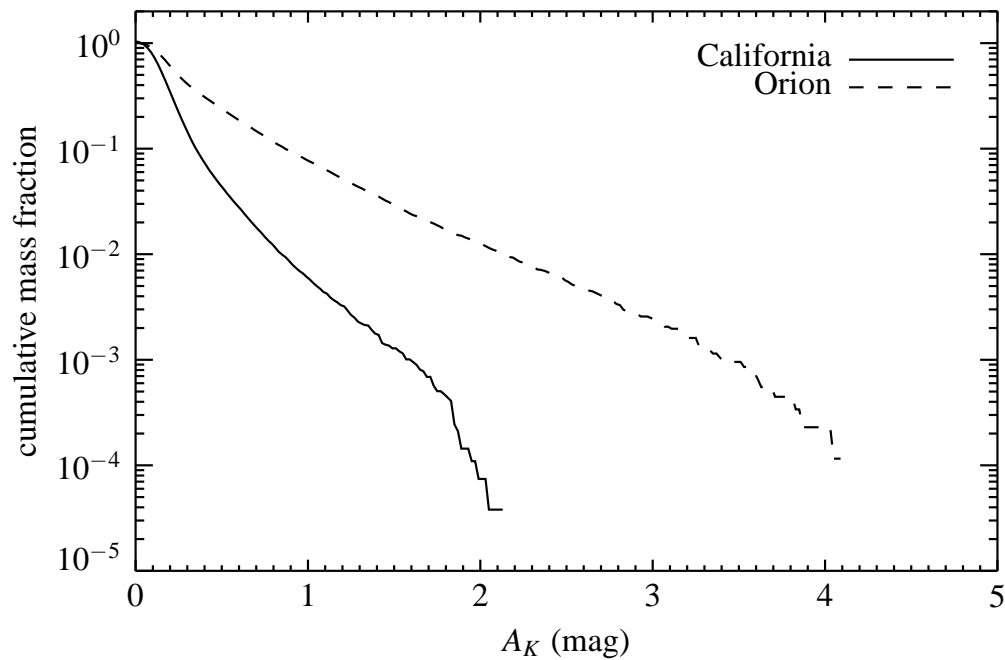}
\caption{The cumulative mass fraction profile of the California Molecular Cloud compared to that of the Orion Molecular Cloud. The measurements of the California cloud were corrected for distance so that the measurements of the two clouds were made at the same physical resolution and properly adjusted for distance dependent foreground and background stellar densities. The two clouds have strikingly different profiles, in particular the California cloud contains substantially less of its mass at high extinction than does the Orion cloud.  \label{extprofile}}
\end{figure}

\clearpage

\clearpage

\begin{table}
\begin{center}
\caption{Cloud Masses\label{tbl-1}}
\vskip 0.1in
\begin{tabular}{ccccr}
\tableline\tableline
$A_K$(mag) & \multicolumn{1}{c}{CMC Mass\tablenotemark{a}} (\msun) & \multicolumn{1}{c}{Adjusted Mass\tablenotemark{b}} (\msun) &\multicolumn{1}{c}{ OMC Mass\tablenotemark{c}} (\msun)& \\
\tableline
$>$ 0.1 &$1.12 \times 10^5$ &$1.09 \times 10^5$  &$7.66 \times 10^4$ \\
$>$ 0.2 &$5.34 \times 10^4$ &$5.03 \times 10^4$  &$5.32 \times 10^4$ \\
$>$ 1.0 &$1.09 \times 10^3$ &$8.12 \times 10^2$  &$6.59 \times 10^3$ \\

\tableline
\end{tabular}
\tablenotetext{a}{for 450 pc distance} 
\tablenotetext{b}{To facilitate comparison the individual map pixels of the CMC 
are here adjusted to have the same  foreground \& background stellar densities and physical resolution as the pixels in the extinction map of the OMC (see text).}
 \tablenotetext{c}{for 390 pc distance}

\end{center}
\end{table}

\begin{deluxetable}{lrrrrcrr}
\tablewidth{0pt}
\tablecaption{Candidate IRAS YSOs}
\tablehead{
\colhead{IRAS Source}           & \colhead{$F_{12}$}      &
\colhead{$F_{25}$}          & \colhead{$F_{60}$}  &
\colhead{$F_{100}$}          & \colhead{Notes}    &
\colhead{ref}}  

\startdata
03507+3801& 0.78&1.62&4.81&10.5& IR RNe, HH 462& 4\\
03530+4120& 0.25&0.71&1.86&3.9& PNe?& 3\\
04056+4011& 0.34&1.03&4.07&4.7&\nodata & \nodata \\	
04067+3954& 0.76&8.25&38.1&82.8& IR RNe &4 \\
04073+3800& 6.48&20.1&47.9&61.5& HH 464; FU*\tablenotemark{d}; IR RNe  &4,7, \\
04088+3834& 0.25&0.33&2.54&4.8&\nodata &\nodata \\
04182+3727& 0.40&0.50\tablenotemark{\dag}&4.35&22.1&\nodata &\nodata\\
04223+3700& 0.71& 1.11& 3.29& 18.8&\nodata & \nodata  \\
04253+3618& 0.35& 1.24& 6.16& 12.2& \nodata& \ \nodata \\
04256+3624& 0.25&  1.03& 3.89& 12.2& \nodata& \nodata \\ 
04265+3605& 0.26&  0.58& 0.78& 11.0& \nodata& \nodata \\
04269+3510& 362.0& 340.0& 3120.0& 5020.0&  LkH$\alpha$101, NGC 1579& 1, 2\\
04269+3550&  2.83&  9.92& 21.3& 15.6& PNe?\tablenotemark{a}& 3 \\
04271+3538& 0.34&  1.75& 5.74&  5.2& \nodata& \nodata \\
04272+3529& 0.26&  0.63& 1.61&  12.8& \nodata& \nodata\\
04273+3548& 0.25&  0.66& 2.64&  15.6& \nodata& \nodata \\ 
04274+3553& 0.36& 1.58& 2.84&   10.6&  PNe?\tablenotemark{a}& 3 \\
04275+3519& 1.58& 1.69& 20.9&   65.1& RNe& 5\\
04275+3531& 0.25& 0.88& 5.36&   6.6& IR RNe\tablenotemark{b}&  4 \\
04275+3452& 0.25& 0.69& 1.47&   9.2& PNe?& 3 \\
04276+3732& 0.33&0.94&4.46&4.9&\nodata & \nodata\\
04315+3617& 0.71&1.37&1.98&6.6&\nodata & \nodata \\
04316+3427& 0.25&0.60&2.92&4.3&\nodata & \nodata\\
04332+3658& 8.68&4.01&0.84&8.3& IRC 40094\tablenotemark{c}& 6 \\

\enddata
\tablenotetext{a}{PNe $=$ Planetary Nebula.}
\tablenotetext{b}{RNe $=$ Reflection Nebula.}
\tablenotetext{c}{Field Giant, SpT: M9.}
\tablenotetext{d}{FU Ori object}
\tablenotetext{\dag}{Poor quality IRAS flux}
\tablerefs{
(1) Andrews \& Wolk 2008; (2) Herbig et al. 2004; (3) Preite-Martinez 1988;  (4) Connelley et al. 2007; (5) Magakian, T.Y. 2003; (6) Vogt 1973; (7) Sandell \& Aspin 1998}
\end{deluxetable}



\begin{thebibliography}{}

\bibitem[Allen \& Davis(2008)]{ad08} Allen, L.~E., \& Davis, C.~J.\ 2008, in Handbook of Star Forming Regions, Volume 1: The Northern Sky, ed. B. Reipurth, (Astronomical Society of the Pacific: San Francisco), p.621,  

\bibitem[Alves et al.(1998)]{allkp98} Alves, J., Lada, C.~J., 
Lada, E.~A., Kenyon, S.~J., \& Phelps, R.\ 1998, \apj, 506, 292 

\bibitem[Andrews \& Wolk(2008)]{aw08} Andrews, S. M. \& Wolk, S. J. 2008, in Handbook of Star Forming Regions Volume 1: The Northern Sky, ed. B. Reipurth, (Astronomical Society of the Pacific, San Francisco), p. 390

\bibitem[Bally(2008)]{b08} Bally, J. 2008, in Handbook of Star Forming Regions Volume 1: The Northern Sky, ed. B. Reipurth, (Astronomical Society of the Pacific, San Francisco), p. 459

\bibitem[Bohlin et al.(1978)]{betal78} Bohlin, R.~C., Savage, 
B.~D., \& Drake, J.~F.\ 1978, \apj, 224, 132 



\bibitem[Chen et al.(1993)]{ctf93} Chen, H., Tokunaga, A.~T., 
\& Fukui, Y.\ 1993, \apj, 416, 235 

\bibitem[Connelley et al.(2007)]{cetal07} Connelley, M.~S., 
Reipurth, B., \& Tokunaga, A.~T.\ 2007, \aj, 133, 1528 

\bibitem[Dame et al.(2001)]{detal01} Dame, T.~M., Hartmann, D., 
\& Thaddeus, P.\ 2001, \apj, 547, 792 


\bibitem[Ekl\"of(1959)]{e59} Ekl\"of, O. 1959 Arkiv for Astronomi, 2, 213

\bibitem[Forbirch et al.(2009)]{flmal09} Forbrich, J., Lada, C.J., Muench, A.A., 
Alves, J. \& Lombardi, M. 2009, \apj, in press.


\bibitem[Gao 
\& Solomon(2004)]{gs04} Gao, Y., \& Solomon, P.~M.\ 2004, \apj, 606, 271 



\bibitem[Herbig et al.(2004)]{HAD04} Herbig, G.~H., Andrews, 
S.~M., \& Dahm, S.~E.\ 2004, \aj, 128, 1233 

\bibitem[Indebetouw et al.(2005)]{ietal05} Indebetouw, R., et 
al.\ 2005, \apj, 619, 931 


\bibitem[Johnstone et al.(2004)]{jdk04} Johnstone, D., Di 
Francesco, J., \& Kirk, H.\ 2004, \apjl, 611, L45 

\bibitem[Kleinmann et al.(1994)]{ketal94} Kleinmann, S.~G., et 
al.\ 1994, Experimental Astronomy, 3, 65 


\bibitem[Lada et al.(1999)]{lal99} Lada, C.~J., Alves, J., 
\& Lada, E.~A.\ 1999, \apj, 512, 250 


 
\bibitem[Lada(1992)]{eal92} Lada, E.~A.\ 1992, \apjl, 393, 
L25 

 
\bibitem[Lilley(1955)]{l55} Lilley, A.~E.\ 1955, \apj, 121, 
559 
 
 
\bibitem[Lombardi(2005)]{Lom05} Lombardi, M.\ 2005, \aap, 438, 169 
\bibitem[Lombardi(2009)]{Lom09} Lombardi, M.\ 2009, \aap, 493, 735 
\bibitem[Lombardi et al.(2009a)]{LAL09a} Lombardi, M., Lada, C.~J.\&  Alves, J. 2009, \aap, submitted.

\bibitem[Lombardi et al.(2009b)]{LAL09b} Lombardi, M., Alves, J., \& Lada, C.~J.\ 2009b, \aap, in preparation
\bibitem[Lombardi et al.(2006)]{LAL06} Lombardi, M., Alves, J., \& Lada, C.~J.\ 2006, \aap, 454, 781 

\bibitem[Lombardi et al.(2008)]{LLA08} Lombardi, M., Lada, C.~J., \& Alves, J.\ 2008, \aap, 489, 143 

\bibitem[Magakian(2003)]{2003A&A...399..141M} Magakian, T.~Y.\ 2003, \aap, 399, 141 

\bibitem[Muench et al.(2008)]{metal08} Muench, A., Getman, K., 
Hillenbrand, L.,  \& Preibisch, T.\ 2008, in Handbook of Star Forming Regions, Volume 1: The Northern Sky, ed. B. Reipurth, (Astronomical Society of the Pacific: San Francisco), p.483


\bibitem[Perryman et al.(1997)]{petal97} Perryman, M.~A.~C., et al.\ 1997, \aap, 323, L49 

\bibitem[Peterson \& Megeath(2008)]{pm08} Peterson, D.~E., \& Megeath, S.~T.\ 2008, Handbook of Star Forming Regions, Volume 1: The Northern Sky, ed. B. Reipurth, (Astronomical Society
of the Pacific: San Francisco), p. 590 


\bibitem[Preite-Martinez(1988)]{PM88} Preite-Martinez, A.\ 1988, \aaps, 76, 317 

\bibitem[Robin et al.(2003)]{rrdp03} Robin, A.C., Reyle, C. Derriere, S., \& Picaud, S. 2003 \aap, 409, 523 

\bibitem[Sandell 
\& Aspin(1998)]{1998A&A...333.1016S} Sandell, G., \& Aspin, C.\ 1998, \aap, 333, 1016

\bibitem[Ungerechts \& Thaddeus(1987)]{UT87} Ungerechts, H., \& Thaddeus, P.\ 1987, \apjs, 63, 645 

\bibitem[Vogt(1973)]{1973AJ.....78..389V} Vogt, S.~S.\ 1973, \aj, 78, 389

\bibitem[Wolf(1923)]{Wolf23} Wolf, M.\ 1923, Astronomische Nachrichten, 219, 109 

\bibitem[Wu et al.(2005)]{Wetal05} Wu, J., Evans, N.~J., Gao, Y., Solomon, P.~M., Shirle, Y.~L., \& Vanden Bout, P.~A. 2005, \apjl, 635, L173


\end{thebibliography}
\end{document}